\documentclass[aps,preprint]{revtex4}

\usepackage{bm}% bold math

\usepackage{graphicx}
\usepackage{dcolumn}
\usepackage{amsmath}
\usepackage{epsfig}

\usepackage{amsthm,amsfonts,amscd}             % Some packages to write mathematics.
\usepackage{eucal}                             % Euler fonts
\usepackage{verbatim}                          % Allows quoting source with commands.
\usepackage{makeidx}                           % Package to make an index.
\usepackage{citesort}
\usepackage{url}                               % Allows good typesetting of web URLs.

\begin{document}
\draft

\title{Optical polarizer/isolator based on a rectangular waveguide with helical
grooves}

\author{Gennady Shvets}
\address{The University of Texas at Austin, Department of Physics, Austin TX 78712}

\newcommand{\ba}{\begin{eqnarray}}
\newcommand{\ea}{\end{eqnarray}}
\newcommand{\be}{\begin{equation}}
\newcommand{\ee}{\end{equation}}
\newcommand{\para}{\parallel}

\begin{abstract}
A chirality-based approach to making a one-way waveguide that can
be used as an optical isolator or a polarizer is described. The
waveguide is rectangular, and chirality is introduced by making
slanted rectangular grooves on the waveguide walls. Chirality of
the waveguide manifests as a strong circular dichroism, and is
responsible for transmitting one circular polarization of light
and reflecting the other. Optical isolation of the propagating
circular polarization is accomplished when the chiral waveguide is
placed in front of a non-chiral optical device. Even the crudest
implementations of chirality are shown to exhibit significant
circular dichroism.
\end{abstract}
%PACS: 52.35.Hr, 52.35.Mw,  42.25.Bs,  52.38.Kd

\maketitle

\newpage

\begin{figure}
   \begin{center}
    \includegraphics[height=7cm]{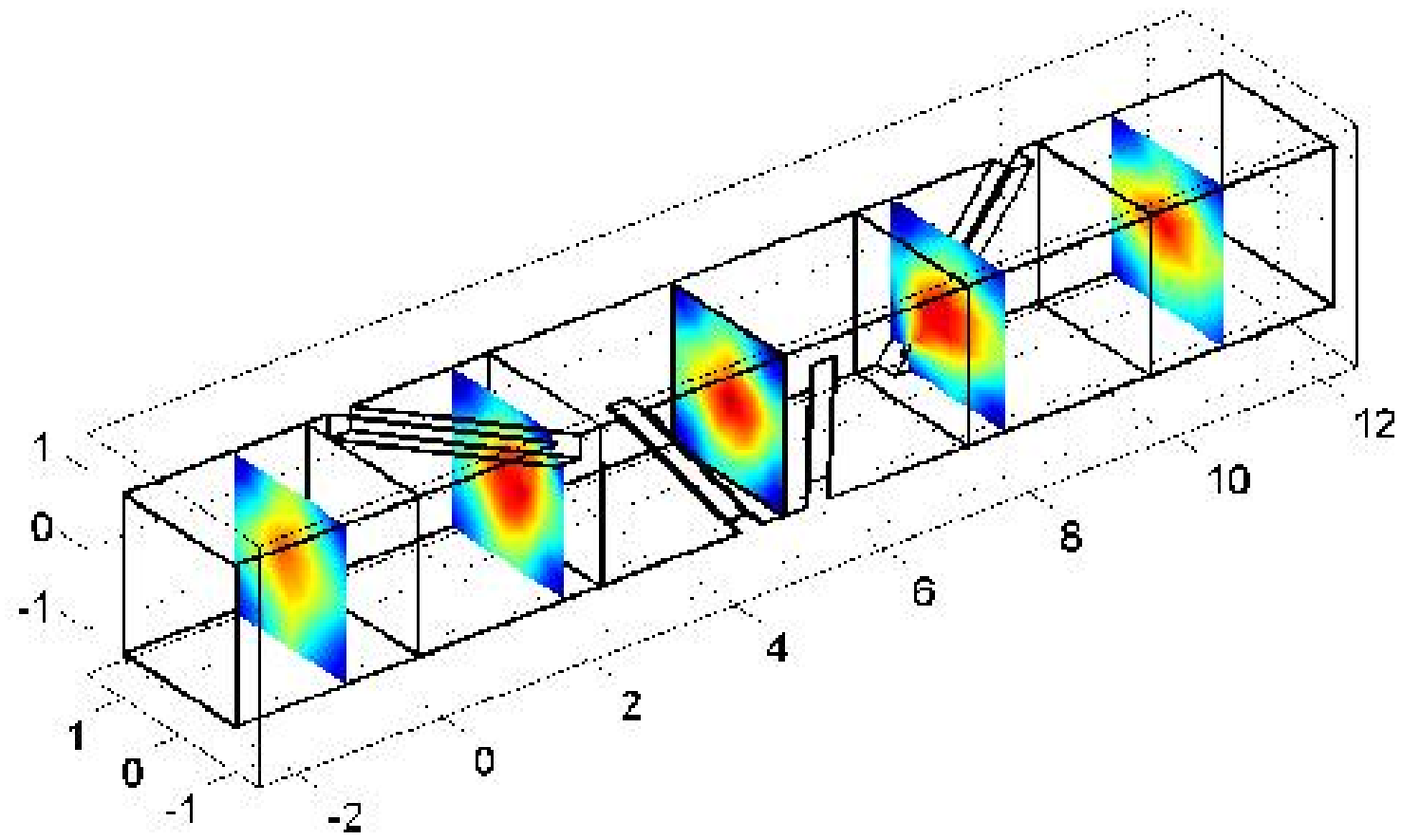}

    \vspace{-10pt}

    \includegraphics[height=7cm]{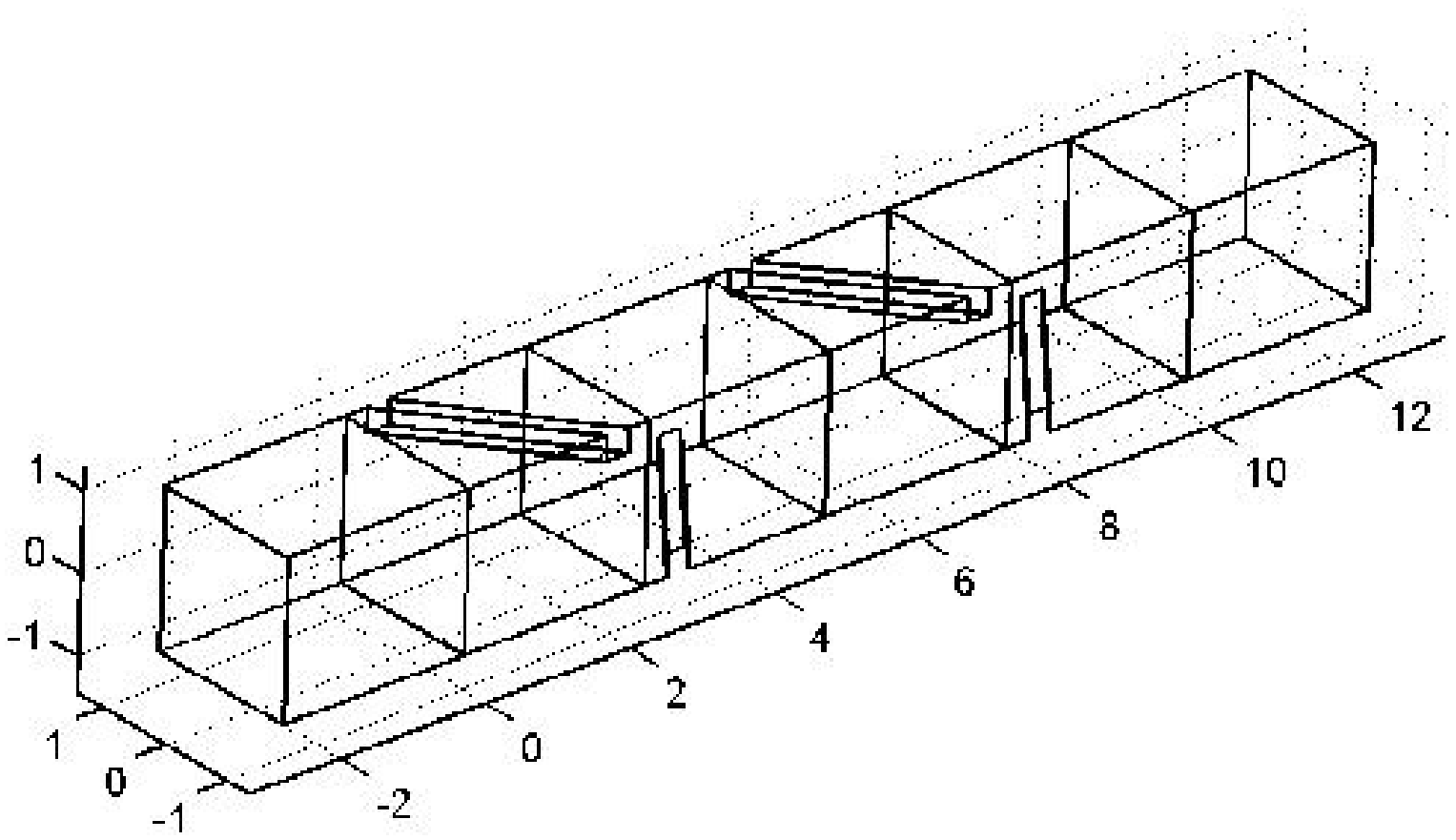}
   \end{center}
    \vspace{10pt}
    \caption{(Color online) Schematic of two rectangular right-handed chiral waveguides
    with helically arranged grooves. (Top): grooves in all four walls.
    Density of the Poynting flux for the injected RHCP wave is
    color coded in several planes to illustrate the preservation
    of the circular polarization for the wave with the same sense
    of rotation as the helical grooves. (Bottom): grooves
    in top and bottom walls. PEC boundary conditions are assumed.
    Distance is normalized to an arbitrary scale $L$ approximately
    equal to a quarter of the vacuum wavelength of the injected
    wave.}
    \label{fig:helical_schematic}
\end{figure}

It is widely believed that the complete integration of electronics
and photonics on a sub-micron scale~\cite{kobrinsky_intercon} must
be accomplished in the near future. Thus the toolbox of integrated
photonics is rapidly expanding, reflecting recent technological
advances in photonic crystals~\cite{vlasov_nature05}, dielectric
waveguides~\cite{vlasov_optexp04}, and magnetooptic
materials~\cite{izuhara_apl00}. Particularly challenging to make
in the integrated form are optical polarizers (devices that
transmit only one light polarization) and related to them
isolators (one-way optical elements that suppress reflection of at
least one polarization). Devices schematically shown in
Fig.~\ref{fig:helical_schematic} solve the problem of developing a
linear one-way optical element by using a rectangular waveguide
with a chiral (arranged as a single right-handed helix)
perturbation to its side walls. Because of the simple rectangular
crossection of the waveguide, and a rather crude implementation of
chirality using periodically arranged slanted grooves in the
waveguide wall, such a device should be relatively easy to
fabricate and integrate with other optical waveguides. As
demonstrated below, propagation of the right- and left-hand
circularly polarized (RHCP and LHCP) laser fields can differ
dramatically: a band of frequencies exists for which only the LHCP
wave propagates through the chiral waveguide (ChW), effectively
making it a simple circular polarizer~\cite{wang_jvac05}.

Chiral twisted fiber gratings with a "perfect" double-helical
perturbation of the refractive index have been suggested as
polarization selective filters in the optical~\cite{kopp_science}
and microwave~\cite{denisov_ieee98,kopp_microwave} frequency
range. Twisting is incompatible with the silicon-based waveguides,
which are also difficult to fabricate with the crossection
different from the rectangular one. The significance of the
proposed structures is that their helicity has a very crude
discrete step and turn symmetry (neither "perfect" nor even
continuous helix) and, therefore, are easy to implement in the
context of integrated optics. Further simplification of the
structure and suppression of Bragg scattering is due to the
single-helix geometry of the grooves.

The proposed chiral optical waveguide can also act as a
polarization-preserving one-way waveguide when inserted between
two optical elements (I and II) that need to be isolated from
reflections. Under a proper choice of the laser frequency
$\omega$, waveguide width $D$, and the helical pitch $\lambda_u
\equiv 2\pi/k_u$, one of the polarizations (e.~g., LHCP) can be
largely transmitted by the ChW when incident from I (that needs to
be isolated) towards II. Let us assume that the non-chiral element
II reflects a small fraction $\eta \ll 1$ of the incident LHCP
radiation. Because the polarization of the reflected radiation is
now RHCP, it will be reflected by the ChW towards II, reflected
again by II as LHCP, and finally emerge from the ChW into element
I. Because two reflections from the element II are involved, the
overall reflection coefficient can be as small as $\eta^{\prime} =
\eta^2 \ll \eta$. Because such isolator is reciprocal, it works
only for one of the two circular polarizations. ChW is thus
similar to another well-known reciprocal optical isolator based on
a quarter wave plate placed behind a linear polarizer, with the
important difference that both the incident on and transmitted
through the ChW electromagnetic waves have the same polarization.
The only practical drawback of a ChW-based isolator is that the
most reflecting elements of the integrated optical network would
have to be operated with the circularly polarized light.

Propagation of electromagnetic waves in a chiral medium
(approximated here by a chiral waveguide) is modelled by the
following
equation~\cite{devries_51,warner_pre01,shvets_tush_pop05}
describing the coupling between the amplitudes $a_{+}$ of the RHCP
and $a_{-}$ of the LHCP components of the electric field:
\begin{eqnarray}\label{eq:couple_1}
&&\left[ \frac{\partial^2}{\partial{x}^2} +
\frac{\omega^2}{c^2}n^2_{+}(x) \right] a_{+} =
\frac{\omega^2}{c^2} g e^{2ik_{u}x} a_{-}, \\ \label{couple_2}
&&\left[ \frac{\partial^2}{\partial{x}^2} + \frac{\omega^2}{c^2}
n^2_{-}(x) \right] a_{-} = \frac{\omega^2}{c^2} g e^{-2ik_{u}x}
a_{+},
\end{eqnarray}
where $n_{\pm}(x)$ are the refractive indices and $g$ is the
strength of the Inter-Helical Bragg Scattering (IHBS). In the
context of wave propagation in the plasma with a helical magnetic
field, Eqs.~(\ref{eq:couple_1},\ref{couple_2}) were shown to
accurately describe coupling between RHCP and LHCP waves through
coupling to a third (idler) plasma wave. As a simple example,
consider the TE$_{01}$ and TE$_{10}$ modes of a square ($-D/2 < y
< D/2$ and $-D/2 < z < D/2$) metallic waveguide propagating in
$x-$direction. RHCP and LHCP modes constructed by linear
superposition have the identical refractive indices $n_{\pm}^2 =
\bar{n}^2(\omega) \equiv 1 - \omega_c^2/\omega^2$, where $\omega_c
= c\pi/D$. Additionally, the two propagation constants will be
modulated with the period $\lambda_u$ due to the realistic
(quasi-helical) perturbation, as will be addressed below by the
first-principles electromagnetic simulations using
FEMLAB~\cite{ref_femlab}. Note that IHBS is a second-order effect:
RHCP wave with $m=+1$ helicity interacts with the helical
perturbation and excites the idler (e.g., TM$_{11}$ with $m=0$)
mode. The idler mode, in turn, interacts with the helical
perturbation and excites the LHCP mode with $m=-1$ helicity. Note
that the identification of RHCP with $m=+1$ mode holds only for
the waves propagating in the $+x$ direction. For the waves
propagating in the $-x$ direction, the $m=+1$ mode corresponds to
the LHCP wave.

To facilitate the qualitative discussion, assume that $n_{\pm}^2 =
\bar{n}^2(\omega)$ does not depend on $z$, i.~e. that the
perturbation is purely helical. Assuming that $a_{+} \propto
\exp{i(k+k_u)x}$ and $a_{-} \propto \exp{i(k-k_u)x}$, a simple
dispersion relation can be derived: $n^2 = n_u^2 + \bar{n}^2 \pm
\sqrt{4\bar{n}^2 n_u^2 + g^2}$, where $n = ck/\omega$ and $n_u =
ck_u/\omega$. Depending on $\omega$, this equation can have zero,
two, or four real roots. It can be analytically shown that,
regardless of the chiral medium parameters $\omega_c$, $k_u$, and
$g$, only two propagating solutions exist for $\omega_1 < \omega <
\omega_2$, where $\omega_{1,2}^2 = (\omega_c^2 + c^2 k_u^2)/(1 \pm
g)$ are the cutoff frequencies. The frequency interval $\omega_1 <
\omega < \omega_2$ is sometimes referred to in the chiral media
literature as the de Vries bandgap~\cite{devries_51,warner_pre01}
for one of the circular polarizations. This remarkable property of
the chiral bandgap enables a polarizer/one-way waveguide based on
the chiral material which transmits only one light polarization
(e.~g., LHCP for the right-handed structure). The approach
described here is to create a {\it reasonable approximation} to a
chiral medium by employing a waveguide with the sidewalls
perturbed in a single helix-like fashion.

As the first example consider a rectangular waveguide shown in
Fig.~\ref{fig:helical_schematic}(a) consisting of four
quarter-wavelength sections with rectangular grooves along the
waveguide walls. Each of the sections is obtained from the
preceding one by translation through the distance $\Delta x =
\lambda_u/4$ and rotation by the angle $\phi = \pi/2$ around the
propagation direction $x$. The wall structure of the waveguide
thus approximates a helical groove while remaining simple and
amenable to standard fabrication techniques: the waveguide itself
and the cuts are rectangular. Although we have assumed, for
computational simplicity, perfect electric conductor (PEC)
boundary conditions at the metal wall, the results are not
expected to be fundamentally different from those for a
high-contrast silicon-based waveguide. Because of the PEC boundary
conditions, the scale length $L$ (approximately equal to a quarter
of the vacuum wavelength) is arbitrary. The waveguide's width and
height (its $y$ and $z$ dimensions, respectively) are $W = H =
2L$, and the pitch of the helix is $\lambda_u = 10 L$. The width
and height of the cuts are $w = h = 0.3L$.

We have numerically solved Maxwells's equations with periodic
boundary conditions at $x=0$ and $x = \lambda_u$ boundaries, and
with PEC boundary conditions at $y=\pm W/2$ and $z = \pm H/2$
boundaries. The waveguide sections $-\lambda_u/4 < x < 0$ and
$\lambda_u < x < 5\lambda_u/4$ shown in
Fig.~\ref{fig:helical_schematic}(a) were not employed in this
source-free (eigenvalue) simulation. The following characteristic
frequencies have been found: $\omega_1 L/c = 1.64$ (lower edge of
the chiral bandgap), and $\omega_2 L/c = 1.70$ (upper edge of the
chiral bandgap). Strong asymmetry between different mode
polarizations propagating is expected inside or near the chiral
bandgap. This property of the ChW was verified by launching RHCP
and LHCP waves through the waveguide structure depicted in
Fig.~\ref{fig:helical_schematic}(a). The forward RHCP and LHCP
waves with the frequencies $\omega = \omega_2$ were launched at
the $x = -\lambda_u/4$. The ratio of the transmission coefficients
(measure of circular dichroism) of the two polarizations is
$T_{R}/T_{L} \approx 0.13$. We have numerically verified the
reciprocality of the structure by launching the two circular
polarizations in the $-x$ direction as well, and obtaining the
same transmission ratio as for the forward waves. Thus, even a
single period of a chiral waveguide acts as a strong polarizer
and, for the LHCP light, a polarization-preserving isolator.

As simple as the ChW shown in Fig.~\ref{fig:helical_schematic}(a)
is, it may still be challenging to fabricate. Specifically, it may
be difficult to create rectangular cuts on all four sidewalls of
the waveguide. Therefore, we have simplified the waveguide
structure even further by making slanted grooves on only two
opposite waveguide walls. Two periods of the structure are shown
in Fig.~\ref{fig:helical_schematic}(b), where the cuts are made on
top and bottom walls. One can still show that this waveguide has a
well-defined helicity with a pitch $\lambda_u = 5L$. However, it
is very crude compared with the idealized helical waveguides
previously considered in the
literature~\cite{denisov_ieee98,kopp_microwave,kopp_science}, and
even with the waveguide shown in
Fig.~\ref{fig:helical_schematic}(a). Nevertheless, the
transmission ratio for the two polarization at $\omega = 1.95
c/L$, or for the same polarizations travelling in opposite
directions is $T_{R}/T_{L} \approx 0.4$. This constitutes a very
strong circular dichroism given that the structure consists of
only two periods. To understand why the transmission of LHCP is so
small, we have plotted the on-axis values of the $m=+1$
(corresponding to forward RHCP and backward LHCP) and $m=-1$
(corresponding to forward LHCP and backward RHCP) components
(dashed and solid lines, respectively) for the incident forward
LHCP (red lines) and RHCP (black lines) waves.

First, consider the incident RHCP wave. The amplitude of the
$m=+1$ component (black dashed line) at the waveguide's exit ($X =
5\lambda_u/2 = 12.5L$) is almost three times smaller than at the
entrance. This is because a significant portion of the forward
travelling RHCP component ($m=+1$) is reflected back into the
$m=-1$ component (black solid line) through the IHBS mechanism.
Therefore, the amplitude of the backwards travelling RHCP
component at the waveguide entrance ($X=-\lambda_u/2=-2.5L$) is
almost equal to that of the incident RHCP wave. There is also
significant conversion into the forward propagating LHCP that is
not described by Eqs.~(\ref{eq:couple_1}) with $n_{+}(x) =
n_{-}(x) \equiv \bar{n}(\omega)$. This conversion occurs due to
the regular Bragg backscattering of the forward RHCP into the
backward LHCP, and the consequent IHBS into the forward LHCP. The
end result is that a strong coupling between the forward and
backward travelling RHCP's results in the low transmission of the
RHCP wave.

Second, consider the incident LHCP wave which has the same sense
of rotation as the chiral groove. The amplitude of its $m=-1$
component (red solid line) at the waveguide's exit is only $25 \%$
smaller than its incident amplitude. This reduction is due to the
usual (non-chiral) Bragg scattering of the forward moving LHCP
wave into the backward moving RHCP. The amplitude of the backward
moving LHCP wave is very small at the waveguide's entrance
implying that there is very little IHBS between the forward and
backward LHCP waves. The above discussion illustrates that there
is a significant asymmetry in IHBS for the LHCP and RHCP waves:
chiral scattering is strong for RHCP and weak for LHCP. It appears
that the resulting circular dichroism can be further enhanced by
controlling the usual (non-chiral) Bragg scattering. This can be
done by introducing additional non-chiral grooves, and by gradual
tapering of the grooves parameters (e.~g.~width) in a multi-period
ChW.

It has also been verified that the chiral nature of the grooves is
necessary for creating circular dichroism of the waveguide.
Specifically, the waveguide cuts have been arranged in a
non-chiral way by modifying the chiral waveguide shown in
Fig.~\ref{fig:helical_schematic}(b): in the new (non-chiral)
waveguide the grooves are slanted in the {\it same} directions on
the top and bottom walls of the waveguide. Transmission
coefficients of the RHCP and LHCP through the non-chiral waveguide
are identical (to the accuracy of our simulation, which is better
than $1\%$) independently of the propagation direction. Therefore,
only a chiral waveguide can serve as a circular polarizer or a
one-way optical element.

In conclusion, we have demonstrated using first principles
electromagnetic simulations that a crude approximation of a chiral
medium based on a rectangular waveguide perturbed by slanted
grooves can act as a circular polarizer which could also be the
basis for an optical isolator. Numerical results are interpreted
on the basis of a model of an ideal chiral medium. The chiral
waveguide shown in Fig.~\ref{fig:helical_schematic}(b) is an
extremely crude approximation of the chiral medium for the
following reasons: (a) it has different cutoff frequencies for the
$z-$ and $y-$ polarizations; (b) coupling is not only between
counter-propagating waves of the same circular polarization, but
also between those with opposite polarizations; (c) the chiral
perturbation of the waveguide is a very crude approximation of a
helical groove. The fact that even two periods of such a simply
designed chiral waveguide possess a high degree of circular
dichroism suggests that a robust design of a polarization
preserving optical isolator/circular polarizer based on chirality
is possible. Future work will extend these results to more
practically relevant silicon-on-insulator waveguides.

This work was supported by the ARO MURI grant W911NF-04-01-0203
and by the DARPA contract HR0011-05-C-0068. Insightful
conversations with Drs.~N.~I.~Zheludev and Y.~A.~Vlasov are
gratefully acknowledged.

\newpage

%\bibliography{helixbib}

\newpage

\begin{figure}[h]
    \begin{center}
        \includegraphics[height=10cm]{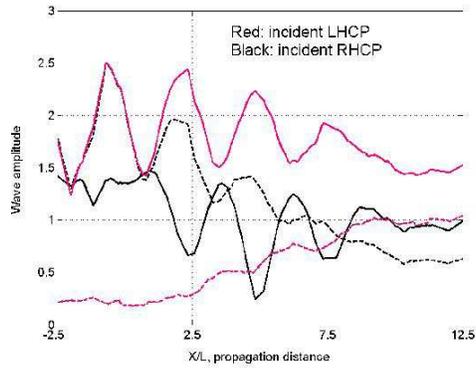}
   \end{center}
    \caption{(Color online) Dashed lines: amplitudes of the $m=+1$ (corresponding
    to forward-moving RHCP and backward-moving LHCP) waves; solid
    lines: amplitudes of the $m=-1$ (corresponding to
    forward-moving LHCP and backward-moving RHCP) waves along the waveguide.
    Two cases are considered: incident RHCP (black lines) and incident LHCP
    (red lines) into a chiral waveguide shown in Fig.~\ref{fig:helical_schematic}(b).
    In the case of incident RHCP wave most of radiation is reflected
    back while almost no reflection is observed for the incident LHCP radiation. The
    overall RHCP transmission is less than half of that of the LHCP.}
    \label{fig:polarizations_twoway}
\end{figure}

\end{document}